\documentclass[12pt]{article}
\usepackage[utf8]{inputenc}
\usepackage[T1]{fontenc}
\usepackage{graphicx}
\usepackage{placeins} 
\usepackage{amsmath}
\usepackage{authblk}
\usepackage{caption}
\usepackage{amssymb}

\usepackage{geometry}
\usepackage{setspace}
\begin{document}

\title{Embedded Direct Ink Writing of Thermoset and Elastomeric Polymers via Frontal Polymerization}

\date{}  % Remove date

% Author: Please give full first and last names for authors and include * after the name of all corresponding authors

\author{
Mohammad Tanver Hossain$^{\text{a,b}}$, 
Yun Seong Kim$^{\text{a,b}}$, 
Pallab Layek$^{\text{b}}$, 
Pranav~Krishnan$^{\text{a,c}}$, 
Youngbum Lee$^{\text{a,c}}$, 
Minjiang Zhu$^{\text{a,d}}$, 
Shubh Singh$^{\text{b}}$, 
Philippe H. Geubelle$^{\text{a,d}}$, 
Paul V. Braun$^{\text{a,b,c}}$, 
Jeffery W. Baur$^{\text{a,d}}$, 
Nancy~R.~Sottos$^{\text{a,c}}$, 
Sameh H. Tawfick$^{\text{a,b}}$, 
and Randy H. Ewoldt$^{\text{a,b,1}}$
}

\date{}
\maketitle

\noindent
$^{\text{a}}$Beckman Institute for Advanced Science and Technology, University of Illinois at Urbana-Champaign, Urbana, IL 61801, USA\\
$^{\text{b}}$Department of Mechanical Science and Engineering, Grainger College of Engineering, University of Illinois at Urbana-Champaign, Urbana, IL 61801, USA\\
$^{\text{c}}$Department of Materials Science and Engineering, Grainger College of Engineering, University of Illinois at Urbana-Champaign, Urbana, IL 61801, USA\\
$^{\text{d}}$Department of Aerospace Engineering, Grainger College of Engineering, University of Illinois at Urbana-Champaign, Urbana, IL 61801, USA\\
$^{\text{1}}$To whom correspondence should be addressed.\\
E-mail: ewoldt@illinois.edu

% Keywords: Please provide a minimum of three and a maximum of seven keywords, separated by commas

%\keywords{FROMP, Embedded 3D printing, Additive manufacturing}

% Abstract should be written in the present tense and impersonal style (i.e., avoid we), and be at most 200 words long
\begin{abstract}
Direct ink writing (DIW) using frontal ring-opening metathesis polymerization offers a compelling route to the rapid and energy-efficient fabrication of thermoset and elastomeric polymer architectures, leveraging a self-propagating exothermic curing reaction. While FP-DIW excels at freestanding path printing due to the rapid solidification, it is constrained by stringent rheological requirements, a lower bound on achievable feature size due to quenching, and the need for the reaction front to closely follow the nozzle during printing. 
Here, we overcome these constraints by leveraging embedded 3D printing to implement FP-DIW with delayed solidification, thereby decoupling shape retention and solidification from ink chemistry and rheology. 
The use of a yield-stress support medium enables extrusion of low-viscosity inks by suppressing gravitational and capillary instabilities, mitigating front quenching at small diameters, and allowing time-delayed solidification to fuse complex, overlapping, and mechanically interlinked features after deposition. 
Two complementary thermal initiation strategies are introduced:\ volumetric dielectric heating via microwaves and surface heating at the boundary of the support bath.
Formulations based on dicyclopentadiene (DCPD), cyclooctadiene (COD), and mixtures thereof, result in tunable final mechanical properties with glass transition temperatures spanning –50 °C to 160°C. 
The versatility of this approach is demonstrated through the fabrication of lattices, springs, mechanically interlocked, and multimaterial architectures.
Compared to printing in air, this embedded approach introduces a substantially broader range of possible formulations, material properties, feature sizes, and architectures.

\end{abstract}

\begin{figure}[!htbp]
\centering
\vspace{-0.1cm}
  \includegraphics[width=0.95\linewidth]{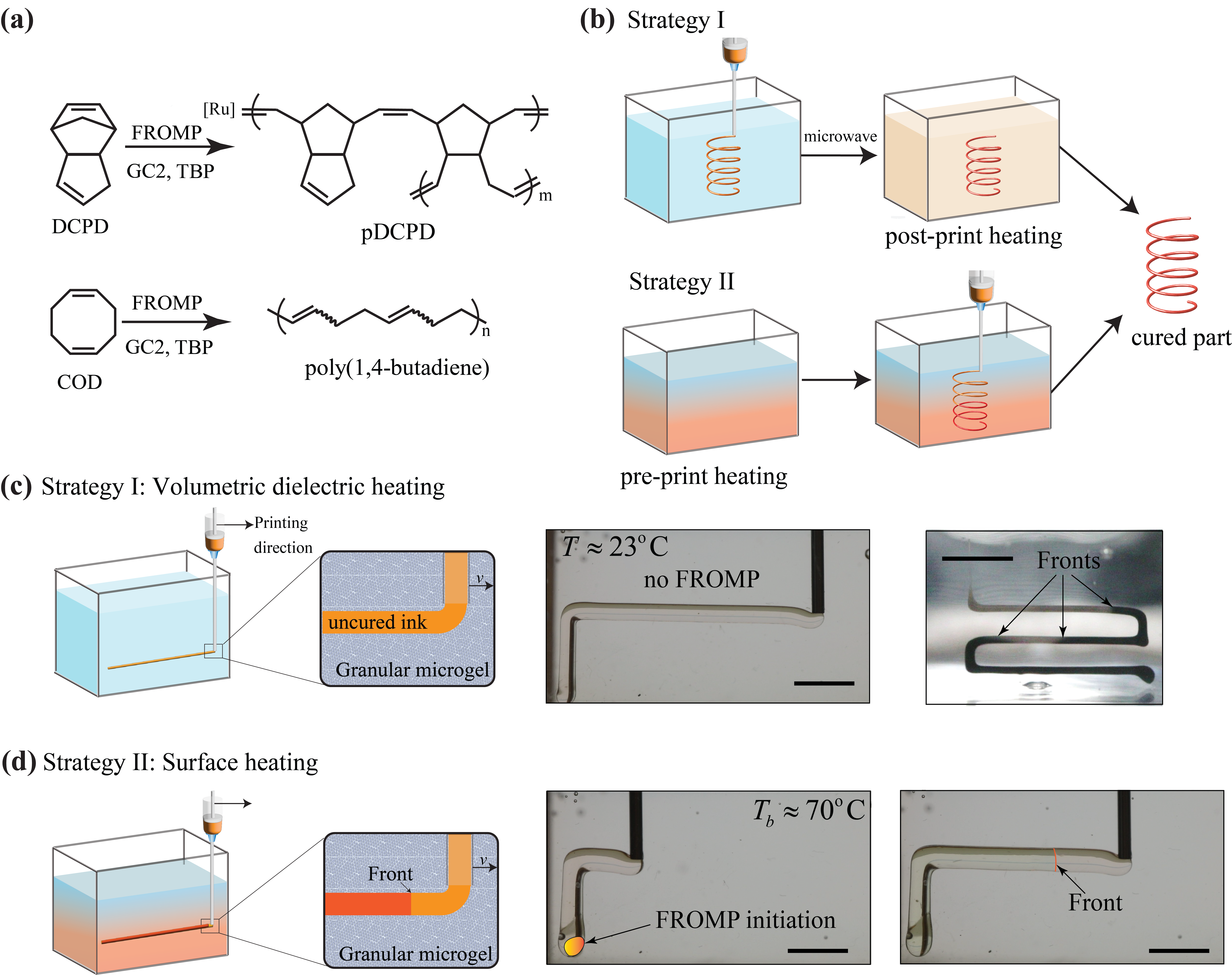}
  \caption{Embedded frontal ring-opening metathesis polymerization (FROMP) using a yield-stress support medium enables extrusion and shape retention of low-viscosity reactive inks.
(a) Reaction schemes for FROMP of dicyclopentadiene (DCPD) and 1,5-cyclooctadiene (COD), catalyzed by second-generation Grubbs catalyst (GC2) and inhibited with tributyl phosphite (TBP), yielding cross-linked poly(dicyclopentadiene) (pDCPD) and linear poly(1,4-butadiene), respectively. (b) Schematic illustration of two embedded curing strategies: Strategy I uses \emph{post-print heating}, where the ink is first printed into a room-temperature support bath and subsequently cured; Strategy II uses \emph{pre-print heating}, where the support bath is heated before printing so that curing is initiated during or immediately after deposition.
(c) Details about strategy I: volumetric dielectric heating. Schematic and optical images showing that FP-based inks printed into an unheated granular microgel support medium ($\sigma_y=28.9$~Pa, $T\approx23\,^{\circ}\mathrm{C}$) remain uncured until microwave exposure triggers post-print frontal polymerization; thermochromic pigment is added to visualize the propagating fronts. 
(d) Details about strategy II: surface heating. Schematic and optical images showing that in a preheated support bath ($\sigma_y=28.9$~Pa, $T_b\approx70\,^{\circ}\mathrm{C}$), thermal initiation occurs within the bath and a visible polymerization front self-propagates through the printed filament; false color is used to visualize the propagating fronts. All scale bars are 10~mm. }
  \label{fig:boat1}
  \vspace{-0cm}
\end{figure}

\section{Introduction}

\quad Additive manufacturing (AM) has transformed how materials and structures are fabricated by enabling the direct, layer-by-layer construction of ageometries~\cite{1,2,3,4,5,6,7,8,9,10}. Among various AM techniques, direct ink writing (DIW) is widely used for printing viscoelastic materials that retain their shape upon extrusion. However, DIW requires the ink to exhibit sufficient yield stress and elasticity to maintain structural fidelity during deposition, which limits the printable formulations, especially for reactive or low-viscosity systems~\cite{1,11,12,13}.

\quad This constraint poses a significant challenge for printing high-performance thermosets such as dicyclopentadiene (DCPD); a monomer that undergoes frontal ring-opening metathesis polymerization (FROMP) to form stiff, cross-linked networks. DIW printing of DCPD has previously leveraged FROMP as a curing strategy, where a localized thermal trigger initiates a self-sustaining exothermic reaction front to convert the extruded resin into solid polymer~\cite{1,14,15,16,17,18,19,20,21,22,23}. This approach of rapid energy-efficient curing, directly after extrusion, enables freeform structures and bypasses the need for conventional thermal post-curing.

\quad Despite its advantages, implementing FROMP in DIW imposes stringent constraints on ink rheology and process parameters. Beyond requiring sufficient viscoelasticity for shape retention, the ink must also support a stable liquid bridge between the nozzle and the advancing polymerization front. If the printing speed exceeds the front velocity (typically 1–2 mm/s~\cite{24,25}), the liquid bridge elongates excessively, resulting in filament sagging, thinning, or breakup. As a result, FP-DIW is often limited to slow print speeds and relatively high-viscosity ($\sim10^2$~Pa.s) inks~\cite{26}. To meet these rheological requirements, DCPD formulations often rely on partial background polymerization to increase viscosity. This introduces added complexity: the resin often requires gelling to reach the target viscosity before printing, and is kept at a low temperature to prevent further background polymerization after reaching the desired viscosity. The challenge becomes even more pronounced for elastomers such as 1,5-cyclooctadiene (COD), which exhibit narrower processing time windows due to rapid bulk polymerization rates~\cite{18,27}.

\quad Embedded 3D printing offers a promising alternative. In this method, inks are extruded into a yield-stress support medium that temporarily holds the deposited features in place during and after printing~\cite{28,29,30,31,32,33,34,35,36,37,38,39,40}. Unlike DIW, embedded printing decouples ink properties from shape retention, allowing the use of low-viscosity formulations that would otherwise be unprintable. This technique has enabled the fabrication of freeform, suspended, or overhanging structures with soft materials such as hydrogels, silicones, and biological inks~\cite{28,32,39}. Extending it to high-performance thermosets would unlock a broader range of structural and functional applications requiring superior mechanical, thermal, and chemical stability.

\quad 
Embedded FP-DIW within a yield-stress support medium provides a fundamentally different processing paradigm for reactive thermoset and elastomeric inks. The support bath stabilizes extruded filaments independent of their intrinsic viscoelasticity, thereby alleviating the stringent rheological constraints imposed by conventional FP-DIW. In this framework, the printing rate is no longer tightly coupled to front propagation speed, and gravitational sagging or diameter-dependent front quenching can be mitigated. We hypothesize that this synergy broadens the practical processing window in multiple dimensions: it enables immediate printing without gelling steps, permits time-delayed solidification initiated on demand, supports low-viscosity and compositionally diverse formulations, allows higher deposition speeds, and facilitates finer feature sizes and direct access to complex geometries with spatially tunable mechanical properties.

\quad Implementing FROMP in an embedded context presents unique challenges. Localized thermal input may be insufficient to sustain polymerization, as heat readily dissipates into the surrounding support bath. Without sustained energy input, the reaction front may quench. A recent strategy by Lee et al.~\cite{41} addressed related limitations in embedded printing of DCPD/COD systems by introducing a chemically-initiated ROMP approach, where a catalyst-rich support matrix acts both as a physical scaffold and as a reservoir of chemical activator, which diffuses into the ink to initiate interfacial polymerization. This approach effectively eliminates the need for external heating, cold storage, or photoinitiation~\cite{41,42}. However, because polymerization begins immediately upon contact with the matrix and proceeds inward via diffusion, this method limits precise control over the timing and location of curing. As a result, premature solidification, liquid-filled cores, or incomplete fusion between overlapping features may arise, particularly in complex or topologically interwoven geometries.

\quad To implement embedded FP-DIW with delayed on-demand solidification, we introduce two complementary thermal initiation strategies: (i) volumetric dielectric heating via microwaves to activate FROMP within a room-temperature support medium; and (ii) bath-heated embedded FROMP, where surface heating at the boundary of a thermally preheated support bath enables front initiation during deposition (Figure~\ref{fig:boat1}a). Both approaches maintain spatial control over curing while preserving the mechanical integrity of the support medium, enabling the direct printing of high-performance thermosets from otherwise unprintable formulations (Figure~\ref{fig:boat1}) and transcending the limitations of chemically initiated FROMP.

\quad
The use of a yield-stress support bath in embedded printing fundamentally changes the rheological design space for reactive inks. While prior work has shown that chemically-initiated ROMP systems can achieve remarkable feature resolution and compositional control~\cite{41}, such approaches polymerize immediately upon deposition, limiting spatial control over solidification. In contrast, our approach provides temporal decoupling between extrusion and curing, allowing for on-demand initiation. However, this added flexibility introduces its own rheological challenges.
Specifically, although the support bath provides structural confinement, if the ink viscosity is too low, instabilities arise, such as spreading around the nozzle tip or break-up during extrusion, leading to poor deposition quality~\cite{35,36}. To probe this boundary, we study the interplay between ink and bath rheology using an effective viscosity ratio framework. As a strategy to modulate ink viscosity while preserving frontal reactivity, we incorporate polybutadiene (PBD) as a rheological additive~\cite{43} to provide ink formulation flexibility without the need for pre-curing, long dwell times, or background viscosity buildup.

\quad Embedded FP-DIW expands the design space for 3D printing thermosets and elastomers by relaxing rheological constraints and enabling delayed solidification, decoupling shape retention and curing from ink chemistry and rheology. In this work, extrusion is performed within a granular microgel–based yield-stress support medium that provides mechanical confinement at rest while remaining fluid-like under shear. This framework allows the use of lower-viscosity, COD-rich formulations that would otherwise be unprintable due to stringent rheological requirements. The microgel support suppresses gravitational and capillary instabilities, mitigates front quenching at small diameters, and enables time-delayed solidification to fuse complex, overlapping, and mechanically interlinked features after deposition. Formulations based on DCPD, COD, and their mixtures offer tunable thermomechanical properties, with glass transition temperatures ranging from –50 °C to 160 °C. Furthermore, we showcase spatial programming of material properties by printing multimaterial architectures with stiff–soft–stiff domains. This versatility supports the fabrication of architected materials, including reentrant lattices, mechanically interlocked chains, and graded domains, and establishes embedded FP-DIW as a general platform for manufacturing high-performance polymer systems with enhanced control over chemistry, structure, and function.

\begin{figure}[t]
\centering
  \includegraphics[width=0.95\linewidth]{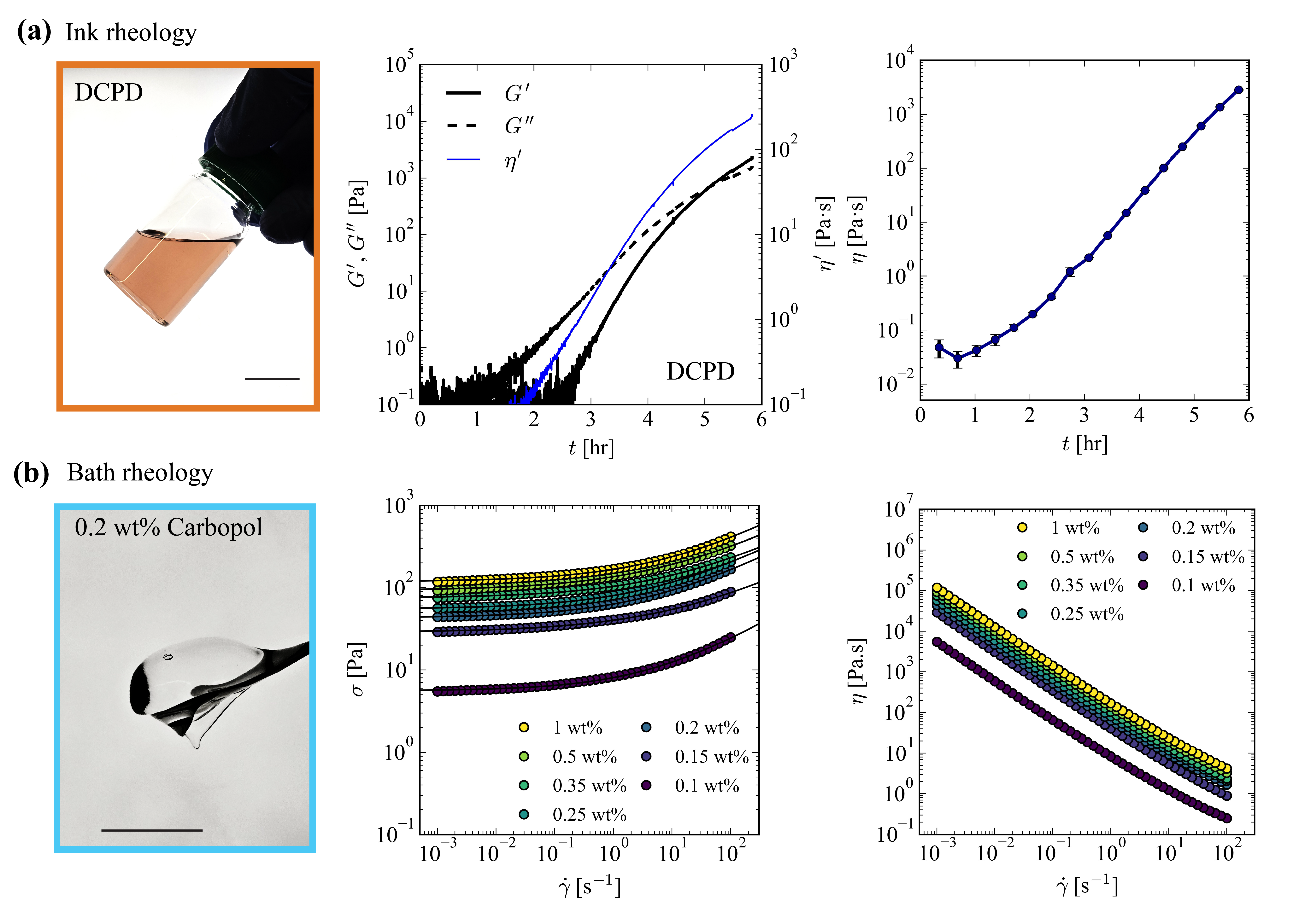}
  \caption{Rheological characterization of DCPD ink and support bath.
  (a) Ink rheology. Left: image of neat DCPD-based ink before curing. Middle: time-sweep oscillatory rheology showing storage modulus $G'$, loss modulus $G''$, and complex viscosity $\eta'$ over time, indicating an increase in viscoelasticity as background polymerization progresses. Right: time-dependent steady shear viscosity at $\dot\gamma=0.1$~s$^{-1}$ showing the increase in ink viscosity, relevant for predicting printability.  
  (b) Bath rheology. Left: image of 0.2~wt\% Carbopol microgel suspension used as a support medium. Middle and right: flow curves for Carbopol suspensions with varying concentrations, showing shear stress and viscosity as functions of shear rate (data fitted to the Herschel-Bulkley model).  
  All scale bars are 10 mm.}
  \label{fig:boat2}
\end{figure}

\section{Results and Discussion}

\subsection{Embedded printing with on-demand frontal polymerization}
\quad Figure~\ref{fig:boat1}a illustrates the reaction schemes for FROMP of DCPD and COD, catalyzed by second-generation Grubbs catalyst (GC2) and inhibited with tributyl phosphite (TBP). The polymerization of DCPD yields cross-linked poly-dicyclopentadiene (pDCPD)~\cite{44}, whereas COD forms linear poly (1,4-butadiene), enabling access to both thermoset and elastomeric materials. We utilize an aqueous granular hydrogel support medium composed of Carbopol microgel particles dispersed in deionized water. Two distinct strategies are used to initiate FROMP within the embedded environment (Figure~\ref{fig:boat1}b). In both cases, ink is extruded into the support medium, where it maintains its shape by the yield stress of the viscoplastic medium. Polymerization is then triggered either through an externally applied thermal trigger or by exploiting the thermal conditions already present in the bath.

\quad In the dielectric heating initiation approach, the ink is extruded into a room-temperature support medium, where it remains uncured during printing (Figure~\ref{fig:boat1}c). After printing, the entire bath is placed in a microwave oven (1000~W), where polymerization initiates after a short exposure, typically within 25~s. The inhibited catalyst is activated by volumetric dielectric heating, triggering frontal polymerization from multiple locations along the printed filament. Thermochromic powder pigment (Atlanta Chem. Eng., Black-Colorless, 70°C) was used for front visualization (Movie SV1). The right image in Figure~\ref{fig:boat1}c shows the characteristic progression of the reaction front through complex geometries, where the cured portion of the filament changes its color from black to colorless. Notably, due to relatively uniform heating of the surrounding bath, the observed front propagation speeds ranged from 1.3–2~mm/s, faster than the front speed observed in open-air DIW at $\sim$1~mm/s~\cite{1,45}. This strategy enables delayed curing, supporting potential applications in multimaterial printing and printing-on-demand workflows~\cite{46}.

\quad Surface heating initiation involves printing directly into a thermally preheated support bath containing a temperature gradient, with the bottom of the container near $T_b\approx$70\,\textdegree{}C (Figure~\ref{fig:boat1}d). Under these conditions, thermal energy within the bath is sufficient to overcome the inhibitor threshold and activate the FROMP reaction as the ink exits the nozzle (Movie~SV2). The resulting polymerization front propagates through the printed path in real time. Optical images reveal a well-defined reaction front, with no external input required beyond thermal bath control. To avoid nozzle clogging (the front may reach the nozzle tip during printing if the print speed is slow), the print speed must be maintained at a minimum of 5--10~mm/s during deposition to keep the nozzle a sufficient distance away from the front. This fast printing speed is particularly suitable for continuous or high-throughput embedded manufacturing processes.

\quad Both methods demonstrate reliable initiation and propagation of frontal polymerization in embedded FP-DIW. Importantly, the printed filament remains stable throughout the reaction due to the localized nature of the exotherm and the controlled rheology of the bath. With volumetric dielectric heating, multiple polymerization fronts are triggered from different locations, as indicated by arrows in Figure~S1, Movie SV3, and Movie SV4. Higher thermal conductivity of the aqueous support medium relative to air~\cite{47} facilitates more uniform heat distribution during initiation. This allows for multiple, spatially distributed initiation sites, reducing the distance any single front must travel and mitigating the risk of quenching due to long propagation lengths. These concurrent fronts eventually impinge and coalesce, forming internal seams that may compromise structural integrity. Prior work has shown that such multi-front interfaces may lead to reduced strain-to-break in the final cured material~\cite{43}. To address this, we implemented a controlled single-point initiation strategy (Figure~S2) where graphite powder is embedded in the gel at the intended initiation point of the printed structure. Owing to its strong microwave coupling as a conductive susceptor, graphite enables localized dielectric heating relative to the surrounding aqueous microgel, promoting rapid, spatially confined thermal activation~\cite{48}. This controlled single-point initiation enables front propagation from a defined origin, propagating at a speed of approximately 1.5 mm/s, which improves curing uniformity and eliminates front collisions (Movie SV5). However, when the volumetric dielectric heating was applied for only 10 seconds, the surrounding gel was not sufficiently heated to support sustained propagation, and the front eventually quenched due to thermal losses to the bath and ambient environment. In surface-heating embedded FROMP, we establish a thermal gradient within the support medium by preheating the base of the bath. Deposition is initiated from this higher temperature zone at the bottom, ensuring that frontal polymerization is preferentially triggered at the base of the printed structure and then propagates upward along the deposition path (Figure~S3 and Movie SV6). This controlled single-point initiation improves front uniformity and minimizes the risk of premature or multidirectional curing. The localized heating supports stable front propagation, with observed speeds ranging from 1.5 to 1.9~mm/s. This approach is particularly effective for maintaining mechanical consistency and print fidelity in vertically extended structures.

\begin{figure}[!htbp]
\centering
  \includegraphics[width=0.9\linewidth]{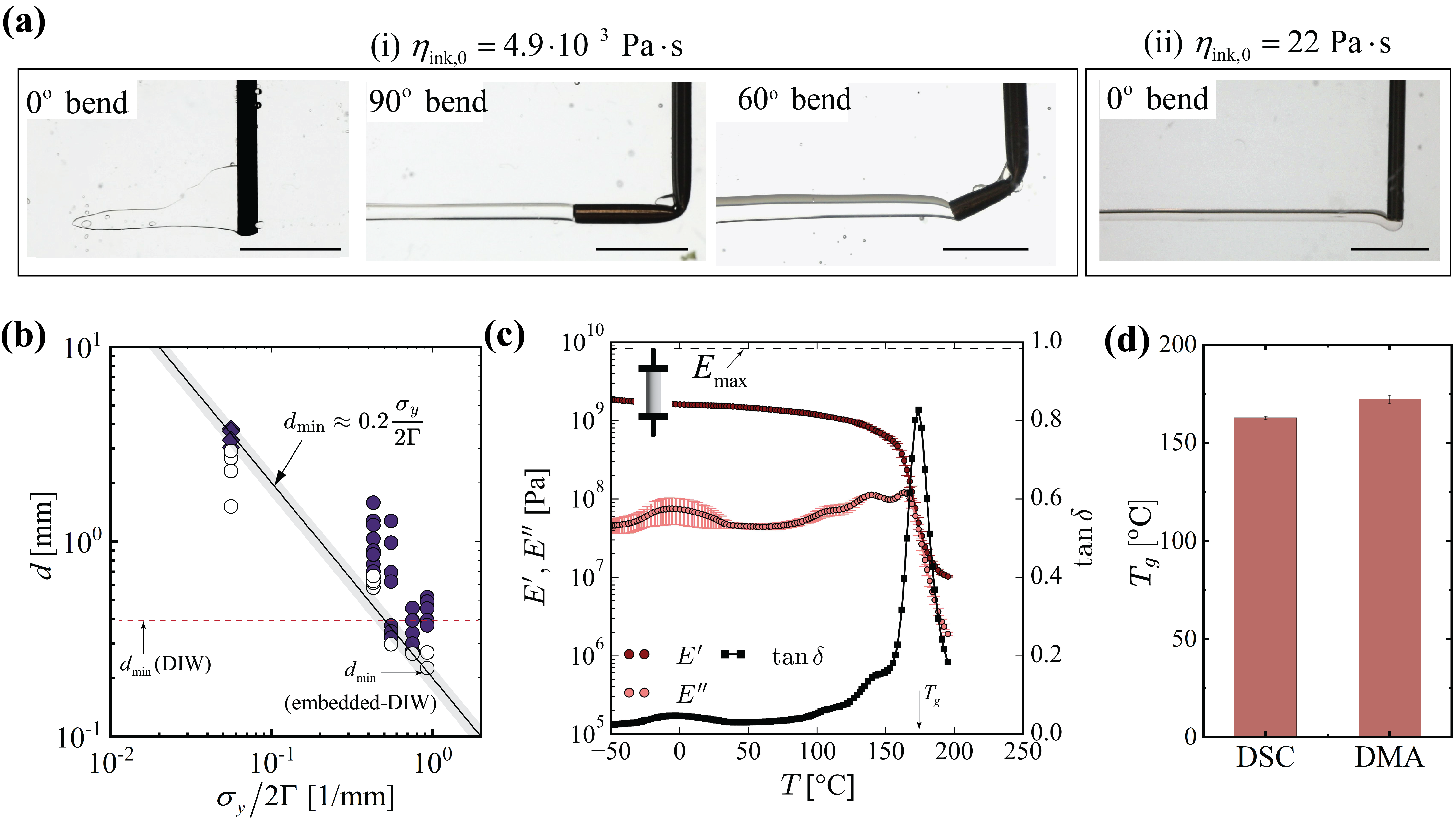}
  %\vspace{-0.5cm}
  \caption{Printing fidelity and thermomechanical properties of pDCPD obtained from embedded printing. (a) Optical images demonstrating print fidelity in embedded 3D printing for different $\eta_{\text{ink,0}}$. For low-viscosity inks ($\eta_{\text{ink},0}/\hat{\eta}_{\text{bath}} \ll 1$), filaments distort significantly when extruded through straight needles. However, sharp-bend needles (90° and 60°) enable successful deposition despite the low viscosity. In contrast, for high-viscosity inks ($\eta_{\text{ink},0}/\hat{\eta}_{\text{bath}}\gtrsim 1$), filaments retain their shape during directional changes. All filaments are printed in 0.2~wt\% Carbopol microgel with yield stress $\sigma_y = 43$~Pa.  (b) Log-log plot of the minimum stable filament diameter $d_{\min}$ as a function of the yield stress to interfacial tension ratio as $\sigma_y / (2\Gamma)$, showing good agreement with the predicted scaling relation $d_{\min} = 0.2\, \sigma_y / (2\Gamma)$ (gray line). Filled symbols indicate filaments that remained stable, while hollow symbols denote broken filaments. (c) Dynamic mechanical analysis (DMA) of a representative pDCPD sample showing storage modulus $E'$, loss modulus $E''$, and $\tan\delta$ as functions of temperature obtained with a clamp fixture with $E_{\max} =8 \cdot 10^9$~Pa.  (d) Comparison of glass transition temperature $T_g$ obtained via differential scanning calorimetry (DSC) and dynamic mechanical analysis (DMA).  All scale bars are 10 mm.}
  \label{fig:boat3}
  %\vspace{-0.5cm}
\end{figure}

\subsection{Rheology of the DCPD ink and aqueous microgel baths}

\quad Successful embedded FROMP printing requires balancing two evolving rheological systems: a time-dependent reactive ink and a viscoplastic support medium. Characterizing both the time-dependent rheology of the ink and the yield-stress behavior of the support medium is essential for assessing printability and filament stability during embedded FROMP. Figure~\ref{fig:boat2}a shows a protorheology image and oscillatory rheology of the DCPD formulation~\cite{49,50,51,52}. As shown in Figure~\ref{fig:boat2}a, the storage modulus $G'(\omega)$ and loss modulus $G''(\omega)$ remained low initially; then increased sharply after approximately 2.5~hr, indicating the onset of network formation and increasing degree of cure. The dynamic viscosity $\eta'(\omega)=G''(\omega)/\omega$ increased by more than 5 orders of magnitude over the 6~hr window, with a rapid increase after the moduli crossover point. The right panel shows the evolution of steady shear viscosity $\eta$ obtained at $\dot{\gamma} = 0.1~\text{s}^{-1}$, which we use as a representative zero-shear viscosity, $\eta_0$, for later analysis. 

\quad In parallel, the support bath must provide sufficient yield stress to confine the extruded filament without impeding nozzle motion. Figure~\ref{fig:boat2}b shows the rheological characterization and protorheology of Carbopol microgel suspensions at different concentrations. All formulations exhibit strong shear-thinning behavior, with high apparent viscosity at low shear rates that decreases over several orders of magnitude with increasing shear rate. The yield stress and the low-shear viscosity increase systematically with Carbopol concentration, offering a tunable platform for the bath medium. The yield stress was extracted by fitting the flow curves to a Herschel--Bulkley model of the form
$\sigma = \sigma_y \left(1 + \left(\dot{\gamma}/{\dot{\gamma}_{\text{crit}}}\right)^n \right)$
where $\sigma_y$, $\dot{\gamma}_{\text{crit}}$, and $ n$ are the fitting parameters~\cite{53}. Fitted parameters are listed in Table~S1. Furthermore, Carbopol suspensions exhibit solid-like linear viscoelastic behavior over the probed frequency range, with \( G' \gg G'' \) and \( \tan\delta < 0.1 \), indicating a predominantly elastic response characteristic of jammed microgel networks (Figure S4). The weak frequency dependence of \( G'' \) further suggests slow structural relaxation and a mechanically arrested microstructure, consistent with yield-stress gel behavior.
 Additionally, temperature-dependent measurements of the 0.25~wt\% Carbopol support gel (Figure~S5) show that while $\sigma_y$ decreases with increasing temperature, it plateaus at approximately 17~Pa above 55$^\text{o}$C. This indicates that the gel retains sufficient yield stress to support printed structures even at elevated temperatures.

\subsection{Print fidelity and thermomechanical properties of pDCPD}

\quad Although embedded printing significantly expands the admissible ink rheology compared to printing in air, filament stability is not unbounded. There remains a lower viscosity limit below which extruded filaments destabilize, spread, or detach despite the presence of a yield-stress support bath (Figure~\ref{fig:boat3}a). Understanding this limit is essential, as it directly governs the printing window, deposition fidelity, and compatibility with low-viscosity reactive resins in embedded FP-DIW.
In general, the fidelity of embedded printing depends critically on the interplay between ink and bath rheology. Prior studies have shown that the effective viscosity ratio, $\eta_{\text{ink}}/\eta_{\text{bath}}$, plays a central role in determining whether printed filaments retain their intended geometry during and after extrusion~\cite{35,36,41}. Here, we systematically investigate this parameter space by varying ink viscosity, bath viscosity, and nozzle angle to identify the stability regime relevant for embedded FP-DIW (Figure~\ref{fig:boat3}a).

\quad DCPD resin (with 5\% 5-ethylidene-2-norbornene, ENB) exhibits a low viscosity of $4.9 \pm 0.13$~mPa$\cdot$s without any background polymerization and behaves as a Newtonian fluid across the relevant shear rate range (Figure~S6). The characteristic shear rate for the ink can be estimated from $\dot{\gamma}_{\text{ink,c}} = 8v_{\text{ink}}/d_i$, the wall shear rate assuming a Newtonian Poiseuille flow in the nozzle, where $v_{\text{ink}}$ and $d_i$ are the average ink velocity and nozzle exit diameter, respectively~\cite{50}.

\quad Estimating a representative viscosity for the support bath is more challenging due to its extreme shear-thinning behavior. Following prior work~\cite{36,50}, we approximate a characteristic shear rate for the bath as $\dot{\gamma}_{\text{bath,c}} = v_{\text{supp}}/d_o$, where $v_{\text{supp}}$ is the bath-relative nozzle speed and $d_o$ is the nozzle outer diameter at the exit. For the conditions in Figure~\ref{fig:boat3}a, nozzle speeds ranged from 5–10~mm/s, with outer diameters of $d_o = 2.1$~mm (for the $0^\circ$ and $90^\circ$ bends) and $d_o = 2.8$~mm (for the $60^\circ$ bend). These values yield bath shear rates of $\dot{\gamma}_{\text{bath,c}} \approx 2$–5~s$^{-1}$, corresponding to a support bath viscosity, $\hat{\eta}_{\text{bath}}=\eta(\dot{\gamma}_{\text{bath,c}}) = 16$–35~Pa$\cdot$s, where $\hat\eta$ denotes the shear viscosity at a characteristic strain rate.

\quad In the absence of any rheological modifier or background polymerization (Figure~\ref{fig:boat3}a(i)), the viscosity ratio is extremely low: $\eta_{\text{ink,0}}/\hat{\eta}_{\text{bath}} \sim 10^{-4}$ where $\eta_{\text{ink,0}}$ is the ink zero-shear viscosity. Under these conditions, filaments extruded from a straight needle deviate significantly from the intended geometry, likely due to spreading or redistribution of the low-viscosity ink along the nozzle surface.

\quad Interestingly, improved filament fidelity is observed even under low viscosity conditions when sharp nozzle bends (60°, 90°) are used. Thus, the deformation dynamics are influenced not only by the viscosity ratio but also by the hydrodynamics of the nozzle motion inside the bath. In particular, interactions between the ink and the nozzle surface, such as wetting and accumulation along the outer nozzle wall in the $0^\circ$ bend case, can destabilize the extruded filament due to continuous shearing against the surrounding bath medium~\cite{54}.

\quad To achieve improved shape fidelity with a straight needle, we deliberately increased the ink viscosity by allowing the DCPD formulation to undergo controlled background polymerization prior to printing. After approximately 4 hours, the steady shear viscosity of the ink increased to $\eta_{\text{ink,0}}=\eta(\dot\gamma = 0.1$~s$^{-1})=$22~Pa$\cdot$s. The actual shear rate experienced by the ink during extrusion likely ranges from 5 to 20~s$^{-1}$ or higher, and DCPD exhibits shear-thinning behavior during ongoing polymerization~\cite{1,19}. We use the shear viscosity at $\dot\gamma = 0.1$~s$^{-1}$ as a representative value, since it approximates the zero-shear viscosity $\eta_{\text{ink,0}}$ and provides a consistent basis for comparison. Under these conditions, printed filaments maintain their geometry even in the challenging $0^\circ$ bend configuration, exhibiting improved shape fidelity and minimal wetting or adhesion to the needle surface (Figure~\ref{fig:boat3}b). 
At this cure state, the viscosity ratio between ink and support bath approaches unity ($\eta_{\text{ink,0}}/\hat{\eta}_{\text{bath}} \sim 1.2$), suggesting that geometric stability is achieved when the effective ink viscosity becomes comparable to that of the surrounding support medium~\cite{36,38,46,55}. This provides a lower-bound (actual $\eta_\text{ink,0}$ is higher) estimate for the stable printing condition shown in Figure~\ref{fig:boat3}a.

\quad While the yield-stress support bath suppresses gravitational sag in embedded DIW printing, capillary forces can destabilize sufficiently thin filaments through breakup. Therefore, the minimum printable filament diameter is determined by the balance between capillary stresses and the yield stress of the surrounding support medium~\cite{28,29,35,41,56}.
Figure~\ref{fig:boat3}b quantifies the influence of bath yield stress and interfacial tension on filament stability. The minimum stable filament diameter, $d_{\min}$, for liquid DCPD without any catalyst is plotted against the ratio of yield stress to interfacial tension, $\sigma_y / (2\Gamma)$, to extract the critical plastocapillary number, also called Yield-Capillary number, $Y_{\Gamma c}$~\cite{29}. The interfacial tension between liquid DCPD and water was measured using the pendant drop method and found to be approximately $33\pm1$~mN/m (Figure~S7). Filaments were printed using a 90$^{\circ}$ bend needle to obtain a circular cross-section (Figure~\ref{fig:boat3}a). Hollow symbols indicate filaments that broke up into droplets due to capillary-driven instability, while filled symbols represent filaments that remained stable after 300~s of observation. The experimental data collapse along a scaling law of the form $d_{\min} = (0.20 \pm 0.04)\, \sigma_y / 2\Gamma$ (gray line), where the prefactor is obtained from fitting the data. This corresponds to a critical Yield-Capillary number of $Y_{\Gamma c} = 0.20$, in agreement with the previously reported value of 0.21 for Newtonian filaments in a microgel granular media~\cite{29}. 
As background polymerization increases the ink viscosity over time, finer filaments become printable, reaching sub-millimeter diameters ($<$1~mm) after approximately 4 hours (Figure~S8). 
To understand the printing limit for filament diameter, we measured the breakup time \( t_{\text{break}} \) of DCPD ink filaments as a function of diameter and ink viscosity in a constant yield-stress support bath, \( \sigma_y = 29~\text{Pa} \) (Figure~S9). As the ink viscosity increases, the breakup time increases significantly. Thus, the ability to print smaller filament diameters may stem from increased viscous resistance to capillary thinning as the ink viscosity rises.

\quad We demonstrate that sub-millimeter features can be printed even in regimes where capillarity would ultimately induce filament breakup. As shown in Figure~S10b, filaments with diameters around 210~\textmu m break up into droplets when printed into a room-temperature bath, for ink with 4 hours of incubation. However, when printed into a heated bath under identical conditions, these filaments remain intact (Figure~S10a), as frontal polymerization solidifies the filament before capillary pinch-off can occur~\cite{28}. This enables the fabrication of finer feature sizes compared to FP-DIW printing in air (Figure~S11 and Figure~\ref{fig:boat3}b). 

\quad Post-cure thermomechanical properties of the pDCPD structures are shown in Figure~\ref{fig:boat3}c–d. Dynamic mechanical analysis (DMA) reveals a high storage modulus, $E' \approx 2 \times 10^9$~Pa at room temperature, indicative of a glassy thermoset (here, $E_{\text{max}}$ indicates the instrument compliance limit~\cite{57}). The glass transition temperature ($T_g$) is marked by a peak in tan$\delta$, and a sharp drop in $E'$, centered around 165\textdegree C. Differential scanning calorimetry (DSC) and DMA measurements yield consistent $T_g$ values (Figure~\ref{fig:boat3}d); the degree of cure was calculated from the residual DSC enthalpy as $\alpha = 1 - H_{\mathrm{residual}}/H_{\mathrm{total}}$, giving $\alpha = 0.99 \pm 0.004$~\cite{27}.
All samples reported in Figure~\ref{fig:boat3}d were cured via volumetric dielectric heating using microwave-assisted initiation.  A representative DSC thermogram of fully cured pDCPD is shown in Figure~S12. The observed values are consistent with prior reports of $T_g \approx 160$\textdegree C for pDCPD prepared by FROMP~\cite{26,27}, demonstrating that the embedded printing strategy successfully preserves the high-performance characteristics of the thermoset.

\begin{figure}[!htbp]
\centering
  \includegraphics[width=0.9\linewidth]{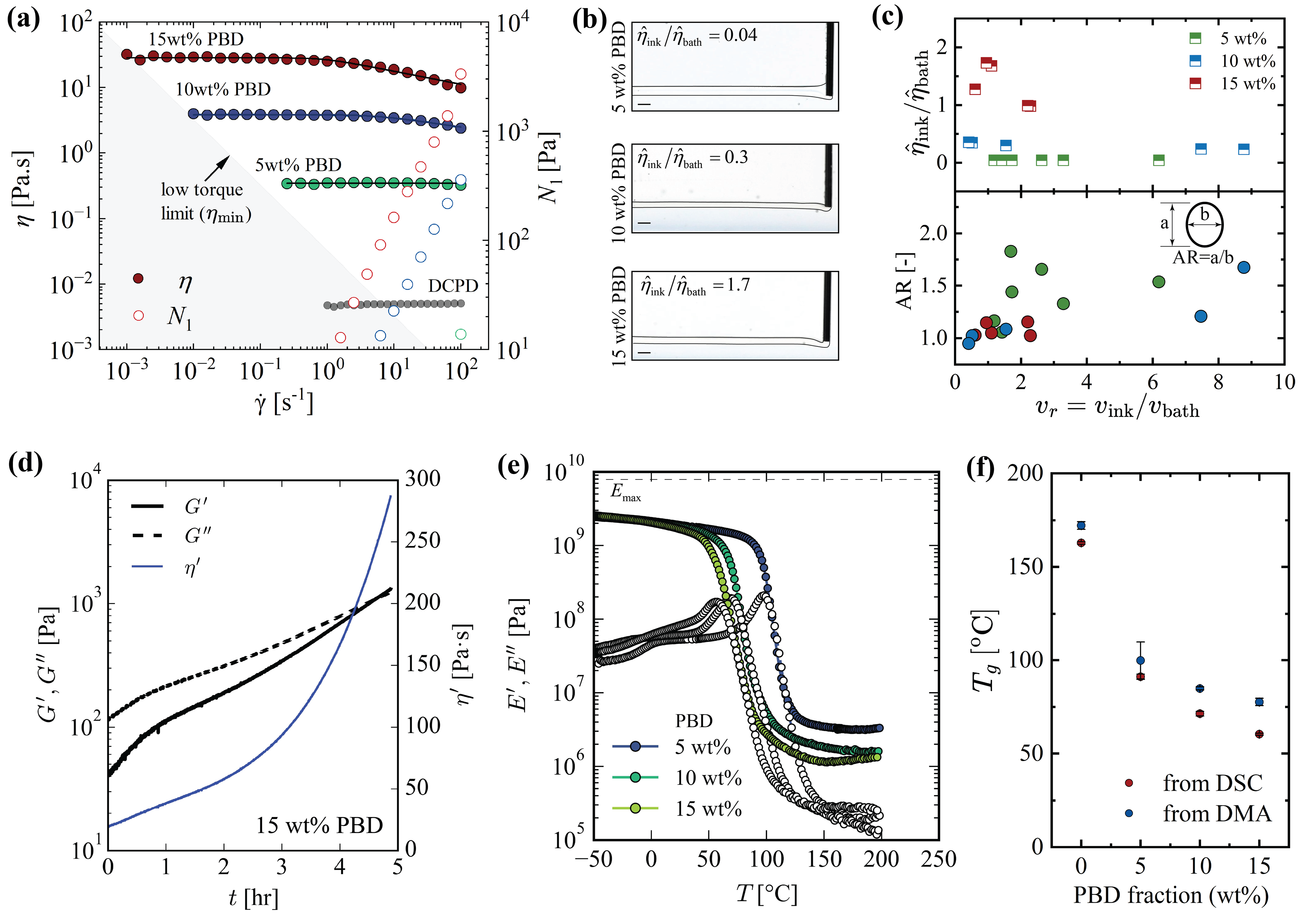}
  \caption{Effect of PBD fraction on ink rheology, print fidelity, and thermomechanical properties of DCPD inks.
  (a) Steady shear viscosity $\eta$ and first normal stress difference $N_1$ of inks with increasing PBD concentration, showing strong shear thinning and elasticity at higher PBD loading. Low-torque limit line (shaded gray area) obtained by using $T_{\min}=0.1~\mu$N.m~\cite{58}.
  (b) Optical images of printed filaments in 0.2~wt\% Carbopol bath at different viscosity ratios $\hat{\eta}_{\text{ink}}/\hat{\eta}_{\text{bath}}$. All scale bars are 1~mm. 
  (c) Top: plot of $\hat{\eta}_{\text{ink}}/\hat{\eta}_{\text{bath}}$ vs. relative printing speed $v_r = v_{\text{ink}}/v_{\text{bath}}$. Bottom: The corresponding aspect ratio (AR) of printed filaments shows fidelity improvement with an increasing viscosity ratio.  
  (d) Time-dependent evolution of viscoelastic moduli ($G'$, $G''$) and complex viscosity $\eta'$ for a 15~wt\% PBD--DCPD ink.  
  (e) DMA characterization of cured materials with increasing PBD content, showing a shift in glass transition.  
  (f) Glass transition temperature $T_g$ measured from DSC and DMA, showing a monotonic decrease in $T_g$ with increasing PBD content. }
  \label{fig:boat4}
\end{figure}

\subsection{Viscosity modified DCPD resins for immediate printing}

\quad  To expand the printability window of DCPD, so that resins are immediately printable rather than needing to wait for background polymerization to reach the required viscosity, we explored the use of polybutadiene (PBD) as a rheology-modifying additive. Resin viscosity was increased while maintaining FROMP reactivity by incorporating 5--15~wt\% PBD (average molecular weight 200{,}000--300{,}000 g/mol) into the DCPD monomer. However, achieving successful frontal polymerization with this highly viscous formulation is non-trivial, as high-molecular-weight PBD is not itself a known FROMP participant and could potentially inhibit front propagation. This formulation has previously been used for printing asymmetric growth structures, and PBD has been shown to co-polymerize with DCPD during FROMP reactions~\cite{43}.

\quad Steady shear measurements for DCPD inks with varying PBD content are shown in Figure~\ref{fig:boat4}a, where increasing PBD concentration systematically raises the ink viscosity and introduces shear normal stress difference $N_1$~\cite{59,60}. At higher shear rates, the inks exhibit non-Newtonian behavior, which is captured using the Carreau model for shear viscosity, 
$\eta(\dot\gamma) = \eta_{\infty} + (\eta_0 - \eta_{\infty}) \left[1 + (K \dot\gamma)^2\right]^{(n-1)/2}$,~where $\eta_0$ is the zero-shear viscosity, $\eta_{\infty}$ is the infinite-shear viscosity, $K$ is the inverse of the critical shear rate, and $n$ is the flow index. Fitted parameters for each formulation are summarized in Table~S2. Based on the model fits, the zero-shear viscosities for 5, 10, and 15~wt\% PBD--DCPD inks were determined to be 0.34~Pa$\cdot$s, 3.84~Pa$\cdot$s, and 28.5~Pa$\cdot$s, respectively. Compared to neat DCPD resin (4.9~mPa$\cdot$s), these values represent viscosity increases of several orders of magnitude. The critical shear rate characterizing the emergence of non-Newtonian behavior is given by $\dot\gamma_{\text{crit}} = 1/K$, and was calculated to be 5~s$^{-1}$ and 0.91~s$^{-1}$ for 10~wt\% and 15~wt\% PBD formulations, respectively. A positive first normal stress difference in shear, $N_1$, is expected to become significant near this shear rate due to the elasticity of polymer chains. Importantly, formulations with $\geq$10~wt\% PBD exhibit significant elasticity, as evidenced by a pronounced $N_1$ around and above $\dot\gamma_{\text{crit}}$. Compared to liquid DCPD resin, different concentrations of PBD resins have observable frequency-dependent viscoelastic properties as shown in Figure~S13. The low-frequency terminal region reveals a viscoelastic relaxation timescale $\tau_0=\lim\limits_{\omega \to 0} G'/(G''\omega)\sim0.5$, 0.04, and 0.005~s for 15, 10, and 5~wt\% PBD loadings, respectively~\cite{43,61}.

\quad Using these viscosified ink formulations, Figure~\ref{fig:boat4}b presents embedded printing results for 5, 10, and 15~wt\% PBD–DCPD inks in a support bath with yield stress $\sigma_y = 42.9$~Pa. These conditions correspond to viscosity ratios $\hat{\eta}_{\text{ink}}/\hat{\eta}_{\text{bath}}$ = 0.04, 0.3, and 1.7, respectively, where $\hat{\eta}_{\text{ink}}$ is the ink viscosity evaluated at the characteristic shear rate $\dot{\gamma}_{\text{ink,c}} = 8v_{\text{ink}}/d_i$. At low $\hat{\eta}_{\text{ink}}/\hat{\eta}_{\text{bath}}$, we observe that the filament may climb or spread along the outer needle surface, whereas higher viscosity formulations maintain straight, well-defined filaments. The quantitative relationship between viscosity ratio ($\hat{\eta}_{\text{ink}}/\hat{\eta}_{\text{bath}}$) and relative extrusion speed ($v_r = v_{\text{ink}} / v_{\text{bath}}$) is shown in Figure~\ref{fig:boat4}c. The corresponding filament aspect ratios (AR$=a/b$), defined as the ratio of the major ($a$) to minor ($b$) axes of the elliptical filament cross section (inset, Figure~\ref{fig:boat4}c), indicate improved print fidelity at higher viscosity ratios.
Notably, 10~wt\% and 15~wt\% PBD formulations can be printed immediately without requiring background polymerization to build up viscosity, while 5~wt\% PBD inks also enable printing without delay, though with slightly reduced shape fidelity (Figure~S14). Even at low concentrations of PBD, viscosity--modified DCPD resins can enhance printability compared to neat DCPD resin.

\quad The curing behavior of a 15wt\% PBD FROMP ink is shown in Figure~\ref{fig:boat4}d. Over a 5-hour window, both the storage modulus ($G'$) and complex viscosity ($\eta'$) increase steadily, indicating that the formulation remains stable over practical printing times. Although the crossover point of $G'$ and $G''$ occurs at approximately the same time as in neat DCPD, the significantly higher initial viscosity of the PBD-modified ink enables immediate printing with improved shape fidelity, eliminating the need for waiting or pre-aging to build viscosity. This combination of long pot life and high initial viscosity makes PBD--modified formulations particularly well-suited for embedded printing using FP-DIW.

\quad Figure~\ref{fig:boat4}e shows DMA results for printed and cured samples with varying PBD content obtained using dielectric-heating initiation. Incorporating PBD compromises the thermomechanical properties of the cured network. Even a small addition, such as 5~wt\% PBD, lowers the glass transition temperature ($T_g$) from approximately 165\textdegree{}C to 100\textdegree{}C, although the storage modulus ($E'$) at room temperature remains largely unchanged. With 15 wt\% PBD, the $T_g$ further decreases to around 65\textdegree{}C (Figure~\ref{fig:boat4}f), which, while significantly lower, is still well above ambient temperature. 
 
\begin{figure}[!htbp]
\centering
\vspace{-0.2mm}
  \includegraphics[width=\linewidth]{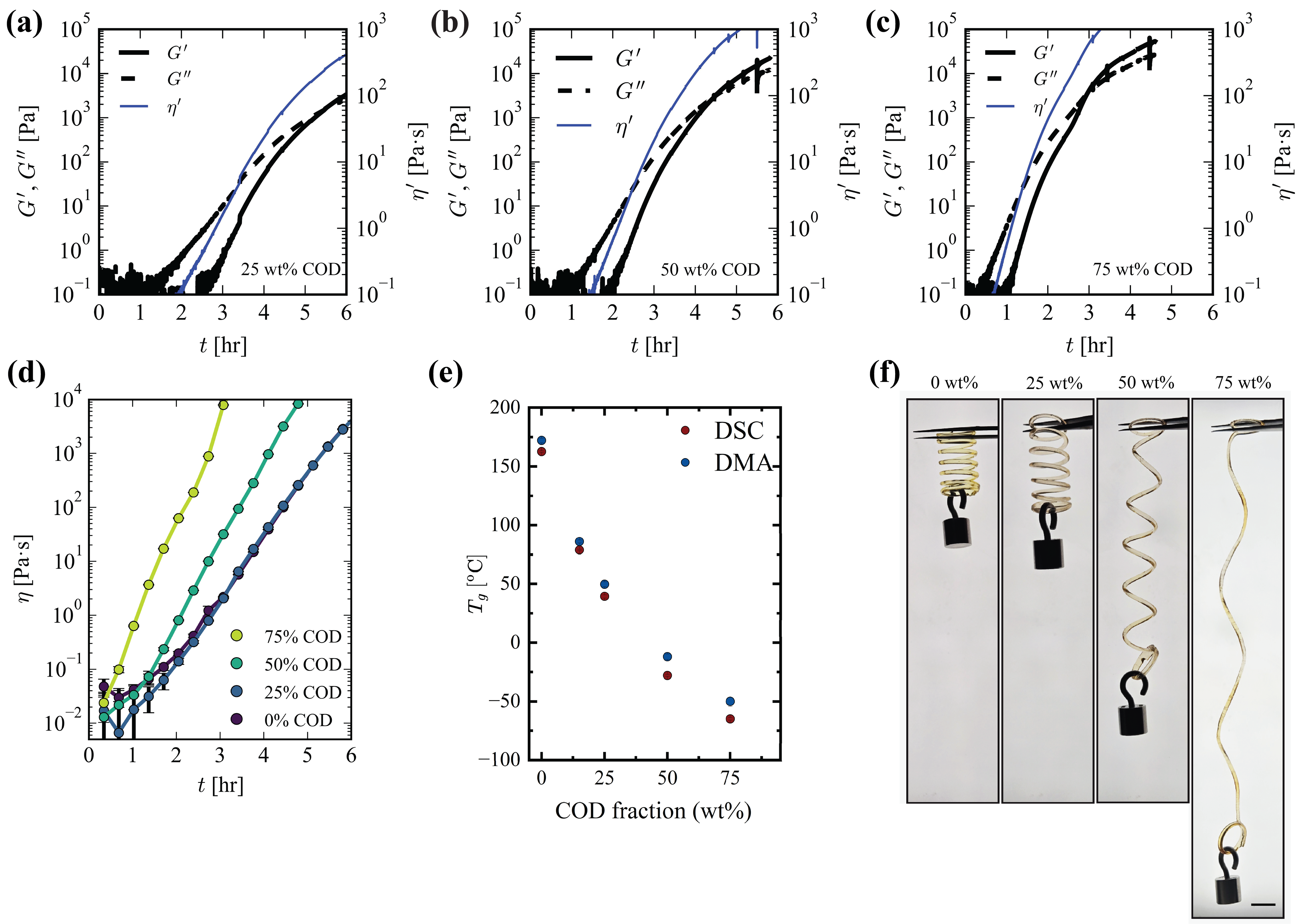}
  \caption{Tuning rheology, thermal transitions, and mechanical behavior of FROMP copolymers by varying COD content.
  (a–c) Time-dependent evolution of storage modulus $G'$, loss modulus $G''$, and dynamic viscosity $\eta'$ for FROMP formulations with (a) 25~wt\% COD, (b) 50~wt\% COD, and (c) 75~wt\% COD at 1 Hz frequency and 0.1\% strain amplitude at 20~$^{\text{o}}$C.  
  (d) Comparison of steady shear viscosity evolution $\eta(t)$ at $\dot\gamma=0.1$~s$^{-1}$ across multiple COD fractions, showing faster viscosity rise with higher COD content.  (e) Glass transition temperature $T_g$ of dielectric-heating FROMP cured materials measured via differential scanning calorimetry (DSC) and dynamic mechanical analysis (DMA), showing a tunable $T_g$ from $\sim$160\textdegree{}C to below $-50$\textdegree{}C as COD content increases.  
  (f) Photographs of printed helical springs composed of copolymers with different COD fractions, each loaded with a fixed weight (10~g), showing a transition from stiff to highly extensible elastomeric behavior. Scale bar is 10~mm.}
  \vspace{-0.5mm}
  \label{fig:boat5}
\end{figure}

\subsection{Synthesis of elastomers and thermosets with tunable thermomechanical properties}

\begin{figure}[!htbp]
\centering
  \includegraphics[width=0.8\linewidth]{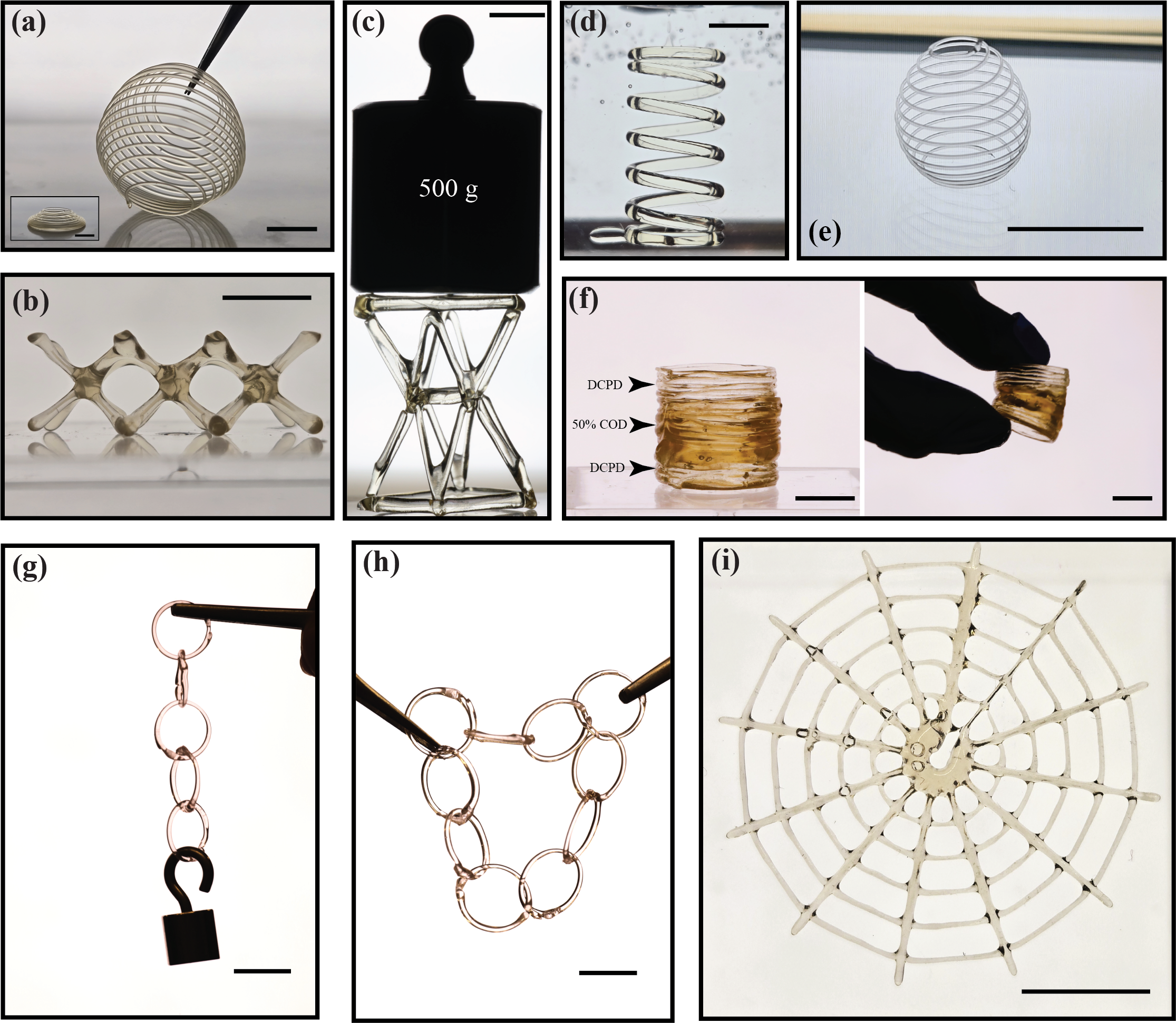}
  \caption{Demonstration of complex 3D structures fabricated via embedded FROMP printing.
  (a) A helical spring geometry with variable helix diameter.
  (b) A side view of a 3×1 body-centered cubic (BCC) unit cell lattice.  
  (c) A truss supporting a 500~g load, illustrating mechanical robustness.  
  (d) A helical spring, still residing in the microgel bath, highlighting smooth vertical path fidelity. 
  (e) A helical spring geometry with filament diameter $d\sim200~\mu$m. (a-e) are printed with neat DCPD resin.
  (f) Multimaterial printing using embedded FROMP to fabricate a hollow cylindrical structure with spatially varying mechanical properties.
  (g) A chain of interlocked rings supporting a metal hook and lock, demonstrating printed mechanical interconnectivity and strength.
  (h) A necklace-like chain of interlocked rings printed in sequence (g-h are printed with 15 wt\% PBD resin)
  (g) A spiderweb-inspired architecture fabricated via continuous filament printing (50 wt\% COD copolymer ink).  
  All scale bars are 10~mm.}
  \label{fig:boat6}
\end{figure}

\quad Comonomer mixtures of COD and DCPD provide a unified chemistry platform capable of spanning over three orders of magnitude in mechanical stiffness, as increasing COD content lowers the glass transition temperature and transitions the material from rigid thermosets to soft, recoverable elastomers~\cite{27}. This compositional tunability offers a powerful route toward architected materials with spatially programmable mechanical properties.
However, implementing COD-rich formulations in conventional FP-DIW is challenging. These systems exhibit more rapid time-dependent viscosity growth during background polymerization than neat DCPD, significantly narrowing the practical printability window and complicating extrusion by rapidly transitioning from low-viscosity liquids to highly viscoelastic states.
Here, we leverage embedded FP-DIW to overcome these limitations and demonstrate printing across a broad compositional range from 0 to 95~wt\% COD. The yield-stress support bath stabilizes low-viscosity COD-rich inks during deposition, while controlled thermal initiation ensures reliable front propagation. Although neat (100~wt\%) COD formulations can slowly crystallize at room temperature, producing opaque and brittle semi-crystalline polymers, incorporation of DCPD suppresses crystallization in COD-rich mixtures, enabling stable elastomeric behavior~\cite{27}.

\quad Figures~\ref{fig:boat5}a–c show oscillatory time sweeps for FROMP formulations containing 25, 50, and 75~wt\% COD. All formulations exhibit a gelation transition as suggested by the crossover of $G'$ and $G''$, which is used here as an approximation of the gelation time among other available criteria~\cite{62}.
The corresponding mutation time, $\tau_{\text{mutation}}=\left|\frac{1}{G'}\frac{dG'}{dt}\right|^{-1}$, decreases from Figure~\ref{fig:boat5}(a) to (c), indicating progressively faster network formation with increasing COD content~\cite{63}. Furthermore, the dynamic viscosity increases more rapidly with increasing COD content, indicating a shorter pot life, defined here as the usable processing window before gelation during which the ink remains sufficiently fluid for extrusion~\cite{26}.

\quad To quantify the printability window of COD–DCPD comonomer mixtures in embedded FP-DIW, we applied the previously defined viscosity ratio, $\hat{\eta}_{\text{ink}}/\hat{\eta}_{\text{bath}}$. Figure~\ref{fig:boat5}d shows the evolution of steady shear viscosity, measured at $\dot\gamma = 0.1\,\mathrm{s^{-1}}$, for formulations containing 0 to 75~wt\% COD. As the COD content increases, the onset of rapid viscosity growth occurs earlier, leading to a progressively shorter pot life.
Importantly, embedded FP-DIW relaxes the viscosity requirements compared to conventional FP-DIW and does not require the ink to reach highly viscoelastic states before printing.
We observe that stable deposition is achieved once the ink attains a moderate viscosity of approximately $\hat\eta_{\text{ink}} \approx 0.5$~Pa$\cdot$s, comparable to that of a 5~wt\% PBD-modified resin. Thus, despite the accelerated viscosity evolution in COD-rich systems, the practical printability window remains accessible with embedded FP-DIW.

\quad For the 95~wt\% COD ink formulation, $G'$ and $\eta$ increase sharply after a short incubation period, signifying fast gelation and a short pot life (Figure~S15). This rapid rheological evolution makes it difficult to maintain stable flow behavior over extended printing durations. We were able to print this formulation within a 20~min time window, during which the viscosity increase is relatively moderate and manageable (Figure~S16). Although direct viscosity measurements during printing are challenging, the spreading behavior near the needle resembles that of 5~wt\% PBD resin (Figure~S16), suggesting a printing viscosity near 0.5~Pa$\cdot$s.

\quad Thermomechanical analysis of the cured samples via DSC and DMA (Figure~\ref{fig:boat5}e and Figure~S17) reveals a systematic decrease in glass transition temperature ($T_g$) and room-temperature stiffness with increasing COD content. The storage modulus $E'(T)$ exhibits a pronounced glassy plateau at low temperature ($E' \sim 10^9$~Pa for low COD fractions), followed by a sharp modulus drop across the glass transition and a rubbery plateau at elevated temperatures (Figure~S17). As COD content increases, both the magnitude of the glassy modulus and the temperature at which the transition occurs decrease progressively. pDCPD (0~wt\% COD) exhibits a $T_g$ near 160~°C with a high glassy modulus, whereas formulations containing 75~wt\% COD display $T_g$ values below –50~°C and rubbery moduli on the order of $10^6$~Pa at room temperature. 
This broad thermomechanical tunability enables a continuous transition from stiff, glassy thermosets to soft, elastomeric networks within a single embedded printing framework. Interestingly, 15 wt\% COD copolymers exhibit comparable $T_g$ values to 15 wt\% PBD-based formulations, with slight variations likely due to the plasticizing effect of phenylcyclohexane (PCH) solvent used to aid the mixing of the catalyst in the PBD--DCPD resin formulation (Figure~S18).

\quad To demonstrate the functional implications of COD fraction, we printed helical springs from inks containing 0–75~wt\% COD and evaluated their deformation under a constant 10~g load (Figure~\ref{fig:boat5}f). Springs printed with 0~wt\% COD remain rigid under load, while those with increasing COD fractions exhibit progressively greater elongation and deformation. This protorheological observation~\cite{50} demonstrates enhanced compliance and deformability with higher COD content.

\subsection{Demonstration of complex 3D structures enabled by embedded FP-DIW}

\quad To demonstrate the versatility and robustness of our embedded FROMP platform, we fabricated a variety of complex 3D structures, including spherical shells, load-bearing trusses, springs, and mechanically interlocked chains, including multi-material (stiff-soft) assemblies (Figure~\ref{fig:boat6}). These examples highlight the method’s capability to produce small, high-fidelity architectures within yield-stress media using formulations that span from rigid thermosets to stretchable elastomers.

\quad Figure~\ref{fig:boat6}a shows a helical spring geometry with variable helix diameter, printed via embedded FP-DIW with DCPD resin. The inset shows the collapsed state at a larger coil diameter, indicating reduced stiffness and increased compliance arising from the small filament diameter. All structures in Figures~\ref{fig:boat6}b–e were also printed using 100\% DCPD ink formulations. Figure~\ref{fig:boat6}b presents a 3×1 body-centered cubic (BCC) unit cell lattice, demonstrating the platform’s ability to fabricate architected structures with controlled porosity and mechanical regularity (see Figure~S19 for the printing sequence). A vertical truss composed of pDCPD is shown in Figure~\ref{fig:boat6}c supporting a 500~g load, underscoring the mechanical robustness of the printed thermoset (see Figure~S20 for the printing sequence). The high fidelity of vertical features and joints confirms the structural integrity achieved during frontal curing. A helical spring, still embedded in the microgel (Figure~\ref{fig:boat6}d), exhibits excellent vertical path resolution. Figure~\ref{fig:boat6}e shows the same helical geometry as in Figure~\ref{fig:boat6}a, but printed with a reduced filament diameter of 210~$\mu$m. Such fine features are not achievable via conventional DIW printing (Figure~S11), as the front would quench due to excessive heat loss and filament slumping due to gravity. In contrast, embedded printing enables stable propagation by leveraging the surrounding bath as a thermal buffer, thereby sustaining the FROMP front.

\quad Figure~\ref{fig:boat6}f demonstrates multimaterial printing using embedded FROMP to fabricate a hollow cylindrical structure with spatially varying mechanical properties. The upper and lower segments are printed with pure DCPD (tensile modulus $E \approx 2$~GPa~\cite{27}), forming stiff thermoset regions, while the central section is composed of a 50~wt\% COD copolymer formulation (tensile modulus $E \approx 7$~MPa~\cite{27}), resulting in a soft, elastomeric domain. This gradient architecture allows the middle segment to bend easily under load while maintaining overall structural cohesion. Such seamless fusion between segments is enabled by delayed, time-controlled initiation in embedded FP-DIW, which permits interfacial bonding before gelation, in contrast to FP-DIW, where rapid curing limits interlayer and intermaterial integration. The protorheology~\cite{50} image further confirms this mechanical contrast: application of force at the stiff ends produces localized bending in the soft middle region, consistent with the large modulus mismatch.

\quad Mechanically interlocked ring structures were directly fabricated within the support medium using 15~wt\% PBD resin, eliminating the need for post-print assembly (Figure~\ref{fig:boat6}g,h). In Figure~\ref{fig:boat6}g, a vertically suspended chain supports a 10 g metal hook. Figure~\ref{fig:boat6}h shows a necklace--style loop structure composed of interlocked rings (see Figure~S21 for the printing sequence).

\quad Finally, a spiderweb-inspired network (Figure~\ref{fig:boat6}i) was fabricated using 50\% COD copolymer ink, illustrating the capacity of embedded FROMP to reproduce biologically inspired architectures with complex filament junctions and radial symmetry.
Together, these structures demonstrate that embedded FROMP printing enables a wide range of structural motifs, spanning rigid and flexible architectures, continuous and interlocked geometries, as well as vertical and suspended elements. The combination of rheological control and thermally triggered curing opens new opportunities for manufacturing architected polymers with both mechanical functionality and geometric complexity.

\section{Conclusion}

\quad This work introduces a robust and versatile platform for embedded 3D printing of thermosets and elastomers using FROMP. By leveraging a yield-stress support medium and decoupling curing from extrusion, we overcome limitations of FP-DIW, including rheological constraints, front quenching, and resolution limits. Two complementary thermal initiation strategies, volumetric dielectric heating via microwaves and bath-heated front activation, enable precise spatial control over solidification while maintaining compatibility with low-viscosity, reactive inks.

\quad Through the incorporation of  PBD and control over DCPD/COD ratios, we independently tune ink rheology and post-cure polymer mechanical properties, accessing a broad processing window with glass transition temperatures ranging from –50 to 160\textdegree C and Young’s moduli spanning over three orders of magnitude. The embedded support bath mitigates gravitational and capillary instabilities, allowing reliable extrusion of inks with low viscosity and delayed curing of topologically complex, interwoven, or suspended features. Notably, the embedded support medium enables a small feature size, conditions that typically lead to front quenching in traditional DIW, likely due to elevated surrounding temperatures and the possibility of multi-front propagation.

\quad The versatility and robustness of this approach are demonstrated through the fabrication of diverse 3D structures, including mechanically interlocked chains, reentrant lattices, compliant springs, and load-bearing trusses, spanning both rigid and elastomeric regimes. Furthermore, we demonstrate spatially programmed architectures with stiffness-graded domains, illustrating the ability to integrate structural and functional complexity from a unified ink platform.

\quad Overall, embedded FP-DIW enables a new manufacturing paradigm that integrates rheological design, reactive processing, and structural control. This strategy opens new pathways for architected polymer systems, with future efforts aimed at expanding multimaterial integration, enabling graded and stimuli-responsive properties, and developing scalable workflows for increasingly complex functional structures.

\section{Materials and methods}
\subsection{Resin preparation}

\quad Solid dicyclopentadiene (DCPD 99, Ultrene) is melted to a liquid state in an oven at 50°C, then mixed with 5~wt\% 5-ethylidene-2-norbornene (ENB, Sigma-Aldrich). The mixed monomers are then filtered through aluminum oxide powders to remove any inhibitors remaining. Antioxidant 4,4'-Methylenebis(2,6-di-tert-butylphenol) (Sigma-Aldrich) is added at 0.5~wt\% to minimize oxidation after the part print. Then the full amount of inhibitor, tributyl phosphite (TBP, Sigma-Aldrich), is added to about 10\% weight of monomer in a nitrogen-filled condition. Grubbs’ second-generation catalyst (GC2, M204, Sigma-Aldrich) is added to the monomer-inhibitor and sonicated for about 1 min. The monomer-catalyst-inhibitor solution is mixed with the remaining monomer (90\%) and stirred for about 2-3 min until the resin is homogenized. The molar ratio is 10,000:1:1 (DCPD:GC2:TBP), matching the formulation reported by Robertson \textit{et al.}~\cite{1} and Kim \textit{et al}.~\cite{43}. Lastly, the resin is transferred to a syringe barrel for printing. Similarly, 1,5-cyclooctadiene (COD, TCI America) is mixed with the DCPD-ENB monomer in different wt\% ratios and filtered through aluminum powder. The same procedure and amount of catalyst-inhibitor are used to prepare the resin.\\

\quad Solvent is required to mix the catalyst with the viscous DCPD-PBD monomer. Polybutadiene (200-300 kDa molecular weight, Sigma-Aldrich) is mixed in various concentrations (5, 10, 15~wt\%) with the monomer and stirred on the hot plate at 90°C with a magnetic stirrer. The batch was typically 200-300~g to finish the mixing process in 24-48~hr. GC2 (M204) is dissolved into phenylcyclohexane solvent (PCH, Sigma-Aldrich), and sonicated for about 2-3~min for uniform dispersion. Then, the inhibitor TBP is added to the catalyst solution in a nitrogen environment. An equal amount of catalyst-inhibitor solution is added to the monomer to prepare the DCPD-PBD resin. The viscous resin is stirred for 2~min and degassed for 3-5~min to removebubbles introduced during the mixing step.~\cite{43}  \\

\subsection{Bath preparation}

\quad Suspensions of aqueous microgel particles were used as yield-stress support media. These granular gel systems consist of cross-linked poly(acrylic acid) microgels that swell upon neutralization, forming a repulsion-dominated structure~\cite{53,64}. Carbopol 980 (Lubrizol, USA) microgel particle powders were dispersed in deionized water at specified concentrations (0.1–1~wt\%) and mixed overnight at 300 rpm. To adjust to pH~7, where swelling and yield stress are maximized, 1 M NaOH solution was added with continued stirring at 200 rpm. Final suspensions were degassed using a planetary centrifugal mixer (THINKY) before experiments.

\subsection{Material characterization}

\quad Rheological characterizations were performed using multiple rheometers and geometries suited to the sample type. Frequency sweep measurements of the printing inks in Figure~\ref{fig:boat2},\ref{fig:boat4} and \ref{fig:boat5} were performed on a DHR-3 rheometer (TA Instruments) using a 25 mm disposable plate with a 1 mm gap, at 1 Hz frequency and 0.1\% strain at 20$^{\text{o}}$. Steady shear data for Carbopol microgel suspensions were obtained using an MCR 702 rheometer (Anton Paar) equipped with a 25~mm parallel plate geometry and 1~mm gap. Both steady shear and oscillatory steady shear data for modified DCPD-based resins were collected using an ARES-G2 rheometer (TA Instruments) with a 25~mm cone-and-plate geometry. The viscosity of liquid DCPD was measured using the same ARES-G2 instrument but with a double-walled concentric cylinders (28~mm inside cup diameter, 29~mm inside bob diameter, 32~mm outside bob diameter, and 34~mm outside cup diameter). Dynamic mechanical analysis (DMA) was performed on cylindrical samples using a film clamp on a DMA 850 (TA Instruments). Samples were 7.5–8.5 mm in length and 1.2–1.6 mm in diameter. Measurements were conducted at a frequency of 1 Hz and a strain amplitude of 0.1\%. The instrument compliance limit, which is associated with a maximum Young’s modulus \( E_{\text{max}} \) that can be measured with a specified geometry, was measured using the relation
$E_{\text{max}} = K_x L/A$
where \( K_x \) is the axial stiffness (in N/m), \( L \) is the sample length, and \( A \) is the cross-sectional area~\cite{57}. Using a nominal value of \( K_x^{-1} = 0.5 \, \mu\text{m/N} \), we obtain $E_{\text{max}} \approx 8 \times 10^9 \, \text{Pa}.$
The glass transition temperatures ($T_g$) of the cured samples were measured via Differential Scanning Calorimetry (DSC) using a TA Instruments DSC250 equipped with an RCS120 chiller. Approximately 5-10~mg of each sample was sealed in a TZero aluminum pan with a Tzero lid to allow nitrogen flow and prevent oxidation during thermal analysis. Samples were subjected to a heating ramp from $-100\,^\circ$C to $200\,^\circ$C at a rate of $10\,^\circ$C/min.

\medskip
\textbf{Data Availability Statement} \par %Please delete the Suppporting Information statement if it is not applicable. Please supply Supporting Information in another file. Supporting information should not be provided in .tex format
The data that support the findings of this study are available from the corresponding author upon reasonable request.

% Acknowledgements
\medskip
\textbf{Acknowledgements} \par %delete if not applicable))
\quad This work was supported as part of the Regenerative Energy-Efficient Manufacturing of Thermoset Polymeric Materials (REMAT), an Energy Frontier Research Center funded by the U.S. Department of Energy, Office of Science, Basic Energy Sciences at the University of Illinois Urbana-Champaign under Award No. DE-SC0023457. R.H.E. acknowledges Anton Paar for providing the MCR 702 rheometer used in some of the rheological experiments.

% References

\end{document}